\documentclass[twocolumn,nofootinbib,showpacs,preprintnumbers,superscriptaddress] {revtex4}

\usepackage{amssymb,amsfonts,amsmath,graphicx,}
\usepackage{color}
\usepackage{enumerate}
\usepackage{braket}
\usepackage{epstopdf}

\usepackage{varioref}
\usepackage{slashed}

\newcommand{\dis}[1]{\begin{equation}\begin{split}#1\end{split}\end{equation}}

\newcommand{\bfrac}[2]{{\left(\frac{#1}{#2} \right)  }}
\newcommand{\eq}[1]{Eq.~(\ref{#1})}

\newcommand\tev{\,{\rm TeV}}
\newcommand\gev{\,{\rm GeV}}
\newcommand\mev{\,{\rm MeV}}
\newcommand\kev{\,{\rm keV}}
\newcommand\ev{\,{\rm eV}}
\newcommand\miliev{\,{\rm meV}}

\newcommand\cm{\,{\rm cm}}

\newcommand\dm{{DM}}
\newcommand\mdm{{m_\dm}}

\begin{document}

\title{Neutrino Oscillations in Dark Matter}

\author{Ki-Young Choi}
\email{kiyoungchoi@skku.edu}
 \affiliation{Department of Physics, BK21 Physics Research Division, Institute of Basic Science, Sungkyunkwan University,  
16419 Korea}

\author{Eung Jin Chun}
\email{ejchun@kias.re.kr}
\affiliation{Korea Institute for Advanced Study, Seoul 02455, Korea}

\author{Jongkuk Kim}
\email{jkkim@kias.re.kr}
\affiliation{Korea Institute for Advanced Study, Seoul 02455, Korea}

\begin{abstract} 
We study neutrino oscillations in a medium of dark matter which generalizes the standard matter effect. 
A general formula is derived to describe the effect of various mediums and their mediators to neutrinos. 
Neutrinos and anti-neutrinos receive opposite contributions from asymmetric distribution 
of (dark) matter and anti-matter, and thus it could appear in precision measurements of neutrino or anti-neutrino oscillations.
 Furthermore, neutrino oscillations can occur from the symmetric dark matter effect even for massless neutrinos. 
\end{abstract}


\preprint{}

\maketitle

{\it Introduction}:
When neutrinos propagate in matter, neutrino oscillations are affected  by their coherent forward elastic scattering 
in which the matter remains unchanged and its effect is described by an effective (matter) potential \cite{Wolfenstein:1977ue}.
This can lead to a dramatic impact in neutrino oscillation, with a resonance enhancement of the effective mixing parameter 
\cite{Mikheev:1986gs}, which is recognized as the MSW effect confirmed as the source of the solar neutrino deficit. 

Recently various medium effects have been considered extensively to study its impact on neutrino oscillations or fit experimental data better in a medium of dark matter (DM)  \cite{Sawyer:1998ac,Hung:2000yg,Berlin:2016woy,Berlin:2016bdv,Krnjaic:2017zlz,Brdar:2017kbt,Liao:2018byh,Davoudiasl:2018hjw,DAmico:2018hgc,Huang:2018cwo,Ge:2018uhz,Capozzi:2017auw,Capozzi:2018bps,Pandey:2018wvh,Nieves:2018vxl,Nieves:2018ewk,Ge:2019tdi,Farzan:2019yvo,Cline:2019seo},  or dark energy \cite{Fardon:2003eh,Kaplan:2004dq,Gu:2005eq,Ando:2009ts,Ciuffoli:2011ji,Klop:2017dim}.
However, there has not been a systematic study of the general medium effect on  neutrino oscillations.

In this article, we derive a general formula describing the medium effect which can be applied to various dark matter models with different mediators to neutrinos. 
The medium effect includes modifications of the neutrino mass and potential which are different for neutrinos and anti-neutrinos 
in a medium of asymmetric dark matter. Remarkably, the neutrino oscillation phenomena can arise solely from the symmetric medium effect in a certain model parameter space.

\medskip

\begin{figure}
\begin{center}
\begin{tabular}{cc} 
 \includegraphics[width=0.2\textwidth]{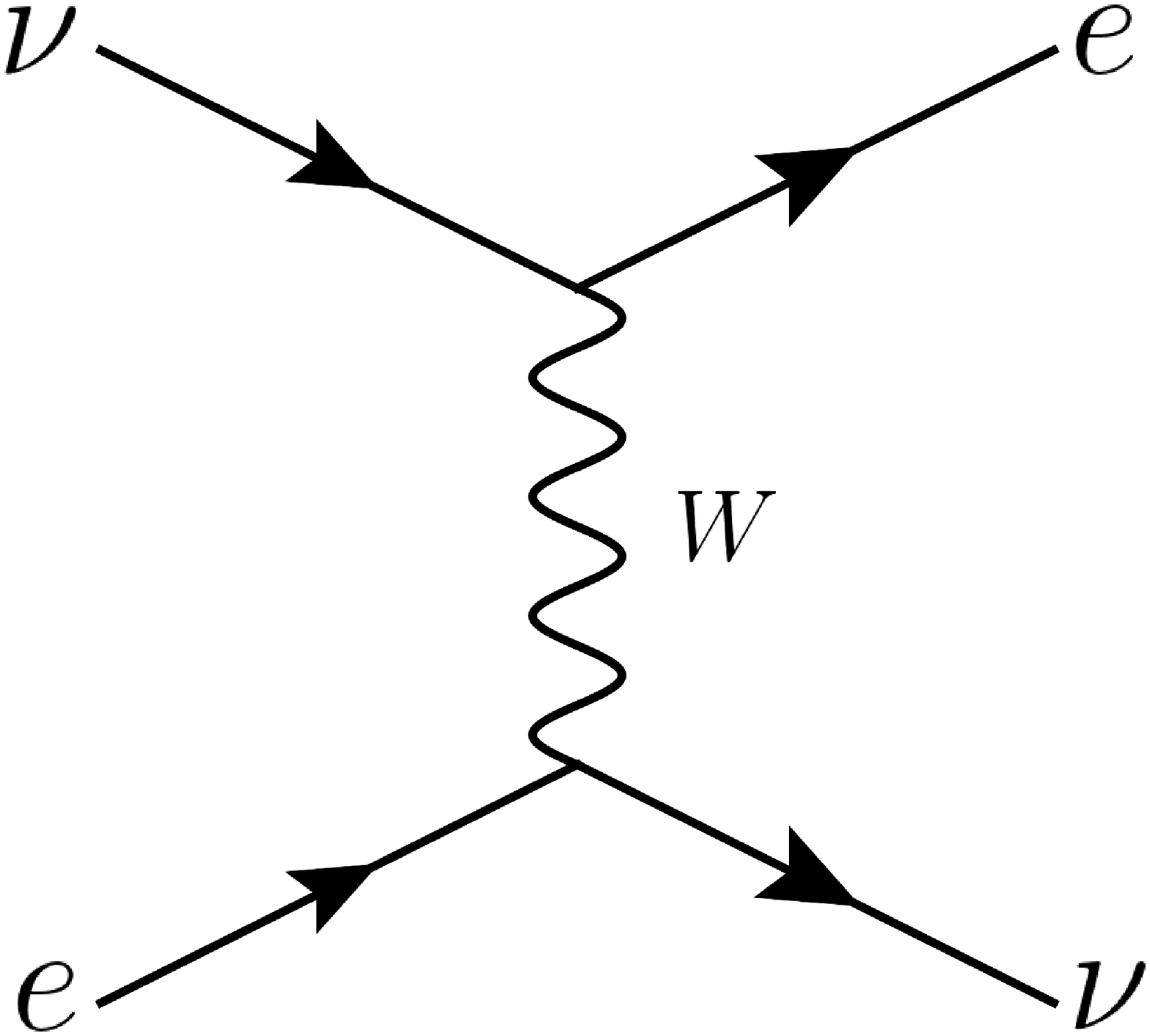}
 &
 \includegraphics[width=0.2\textwidth]{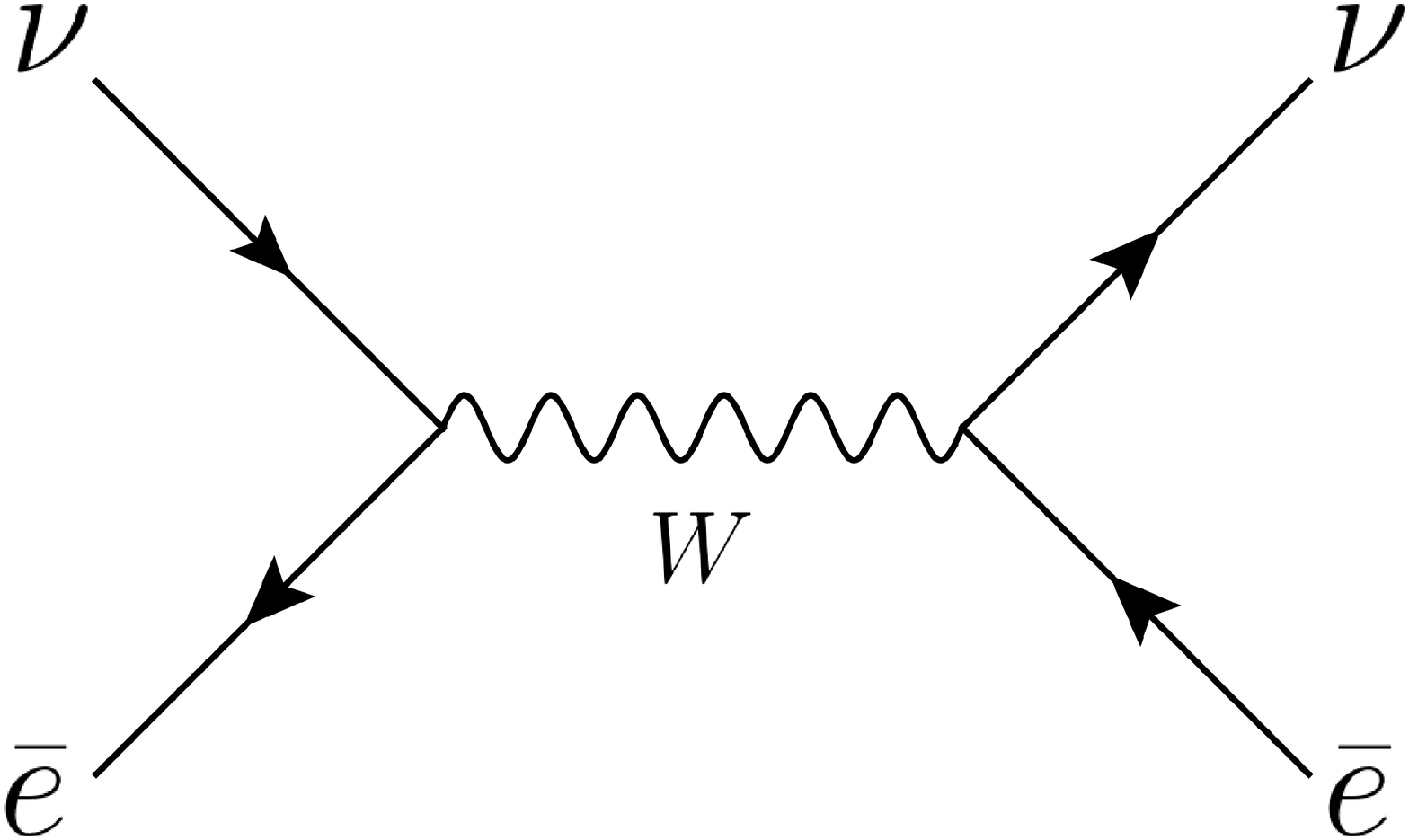}
     \end{tabular}
\end{center}
\caption{ Feynman diagrams for  the  scattering of  neutrino with electron or positron through the $W$ boson exchange in SM.}
\label{diagram_SM}
\end{figure}

{\it Standard matter effect in a medium of electrons and positrons}: \quad 
Let us begin by revisiting the standard calculation of the matter effect driven by the charged current interaction of the Standard Model (SM). For this, we consider neutrino/anti-neutrino propagation in a general background densities of electrons $N_e$ and 
positrons $N_{\bar e}$.  Calculating the coherent forward scattering for the $u$ and $s$ channel processes in Fig.~\ref{diagram_SM}, 
one can find the generalized matter potential:
\begin{equation} \label{VSM}
V^{SM}_{\nu, \bar\nu} = \sqrt{2} G_F (N_e + N_{\bar e}) \frac{ \pm \epsilon\, m_W^4 - 2 m_W^2  m_e E_\nu}{m_W^4 - 4 m_e^2 E_\nu^2},
\end{equation}
where $G_F\equiv g^2/(4\sqrt2 m_W^2)$ is the Fermi constant, and  $\epsilon \equiv (N_e - N_{\bar e})/(N_e + N_{\bar e})$ describes the asymmetry of electron and positron distributions. 
Note that it reduces to the Wolfenstein potential $\pm\sqrt2 G_F N_e$ for neutrinos and anti-neutirnos, respectively in the ordinary situation: $\epsilon=1$ ($N_{\bar e}=0$) and $m_W^2 \gg 2 m_e E_\nu$.  Furthermore, we notice that the matter potential at high energy takes the form:
\begin{equation}
 V^{SM}_{\nu, \bar\nu} \approx \frac{\sqrt{2} G_F m_W^2 (N_e+N_{\bar e})}{2 m_e E_\nu},
\end{equation}
which mimics the standard oscillation parameter $\Delta m^2 /2E_\nu$ acting in the same way for neutrinos and anti-neutrinos. 
The basic formula \eq{VSM} already describes main features of a more general medium effect, which will be described below.

\medskip

{\it Variant models of medium and mediator}:
Staying close to the Standard Model case, we can consider a model of (Dirac) fermionic dark matter $f_i$ and dark photon $X$ as its messenger to neutrinos:
\begin{equation} \label{model1}
{\cal L}_{int} = g_{\alpha i} \bar f_i \gamma_\mu P_L \nu_\alpha X^\mu + h.c.
\end{equation}
Introducing the flavor-dependent couplings, we will get a flavor-dependent medium potential generalized from \eq{VSM} with $m_e= m_{f_i}$ and $m_W=m_X$. 

The second model consists of fermionic dark matter $f$ with Dirac mass $m_f$ and bosonic messenger $\phi_i$ with mass $m_\phi$:
\begin{equation} \label{model2}
{\cal L}_{int} = g_{\alpha i} \bar f  P_L \nu_\alpha \phi_i^* + h.c.
\end{equation}
where we introduced different flavors in the mediator $\phi_i$ instead of the dark matter $f$ for simplicity. 
This model also leads to the same kind of medium potential with $m_e = m_f$ and $m_W=m_{\phi_i}$. 

The third model which is of our particular interest  has complex bosonic dark matter $\phi$ with mass $m_\phi$ and fermionic messenger $f_i$ with Dirac mass $m_f$: 
\begin{equation} \label{model3}
{\cal L}_{int} = g_{\alpha i} \bar f_i  P_L \nu_\alpha \phi^* + h.c.
\end{equation}
which generates the medium potential as well as corrections to the neutrino mass as we will see later.

For all the cases, we will use the unified notations of $m_{DM}$ for the dark matter mass, $\rho_{DM}= m_{DM} (N_{DM}+N_{\overline{DM}})$ for the total dark matter energy density, and 
\begin{equation}
\epsilon\equiv \frac{N_{DM}-N_{\overline{DM}}}{ N_{DM}+N_{\overline{DM}}},
\end{equation}
for the asymmetry between the dark matter and anti-dark matter number densities. 

\medskip

{\it General formulation}:
Neutrino/anti-neutrino propagation in a medium can be described by the following minimal form of the equations of motion in the  momentum space:
\dis{
(\slashed{p} - \slashed{\Sigma}) u_L &= (M^\dagger  + \bar \Sigma_0) u_R,\\
(\slashed{p} - \bar\Sigma \!\!\!\!\slash ) u_R &= (M+ \Sigma_0) u_L, }
 
where $M$ is the symmetric (Majorana) neutrino mass matrix; $ \slashed{\Sigma}\equiv \Sigma_\mu \gamma^\mu$, 
$\bar\Sigma \!\!\!\!\slash \equiv \bar\Sigma_\mu \gamma^\mu$, $\Sigma_0$, and $\bar\Sigma_0$   
are corrections coming from the effect of  coherent forward scattering of neutrinos/anti-neutrinos 
within medium.
Here $\Sigma_\mu$, $\bar\Sigma_\mu$ are hermitian matrices.
Note that we used $u_L$ and $u_R$ to represent the neutrino and anti-neutrino state, respectively. 
Equivalently one may use $v_L$ for the anti-neutrino state using the relation: $\bar u_R = -v_L^T C$ and $u_R=C \bar v_L^T$ where 
$C=-i \gamma^2 \gamma^0$ is the charge-conjugation matrix. 

In a Lorenz invariant medium,  $\Sigma\!\!\!\!\slash$ and $\bar \Sigma\!\!\!\!\slash$  can be expressed  by
\begin{equation}
\Sigma\!\!\!\!\slash = p\!\!\!\slash\, \Sigma_1 + k\!\!\!\slash\, \Sigma_2; ~
\bar\Sigma\!\!\!\!\slash =  p\!\!\!\slash\, \bar\Sigma_1 + k\!\!\!\slash\, \bar\Sigma_2 ,
\end{equation}
where $k$ is the energy-momentum of the dark matter which we will take $(k_0, \vec{k})=(k_0, \vec{0})$  corresponding to averaging over random momentum distribution, and $k_0$ becomes the dark matter mass $m_{DM}$ in the non-relativistic medium. 
The scalar terms  $\Sigma_0/\bar\Sigma_0$ appear in some situations
\cite{Sawyer:1998ac,Hung:2000yg,Berlin:2016woy,Ge:2018uhz,Davoudiasl:2018hjw,DAmico:2018hgc,Capozzi:2018bps}, which will not be discussed further in this article.
\medskip

Recall that the SM matter effect contributes to the vector current terms $\Sigma_2/\bar\Sigma_2$. 
Similar terms are generated in the models in Eqs.~(\ref{model1},\ref{model2},\ref{model3}) and thus a medium potential similar 
to the standard matter potential \eq{VSM} is produced. On the other hand, the correction to the neutrino kinetic term,   $\Sigma_1/\bar\Sigma_1$,  arises only in \eq{model3}. 

The canonical basis of the kinetic term can be recovered by the transformation 
\dis{ \label{uLuR}
 u_L  \simeq & \left(1 + \frac{\Sigma_1}{2}\right) \tilde{u}_L,  \\
 u_R \simeq  & \left(1+ \frac{\bar\Sigma_1}{2} \right)  \tilde{u}_R ,
}  
in the leading order of $\Sigma/\bar\Sigma$.
This leads to  the medium-dressed neutrino mass matrix
\dis{
{\tilde M} \simeq  \left(1+ \frac{\bar\Sigma_1}{2} \right)  M  \left(1 + \frac{\Sigma_1}{2}\right) ,
}  
and thus we obtain
\dis{
(\slashed{p} - \slashed{k} \Sigma_2)\tilde u_L &= \tilde M^\dagger \tilde u_R,\\
(\slashed{p} - \slashed{k} \bar\Sigma_2) \tilde u_R &= \tilde M  \tilde u_L.
}
This takes the same form as in the case of the SM matter effect and thus one obtains  neutrino/anti-neutrino propagation Hamiltonians 
  \begin{eqnarray}
H_\nu &=& E_\nu + {\tilde M^\dagger \tilde M \over 2 E_\nu} + k^0 \Sigma_2, \\
H_{\bar\nu} &=& E_\nu + { \tilde M \tilde M^\dagger \over 2 E_\nu} + k^0 \bar\Sigma_2,
\end{eqnarray}
in the ultra-relativisitc limit: $|\vec{p}_\nu| \approx E_\nu$.

\begin{figure}[!t]
\begin{center}
\begin{tabular}{cc} 
 \includegraphics[width=0.2\textwidth]{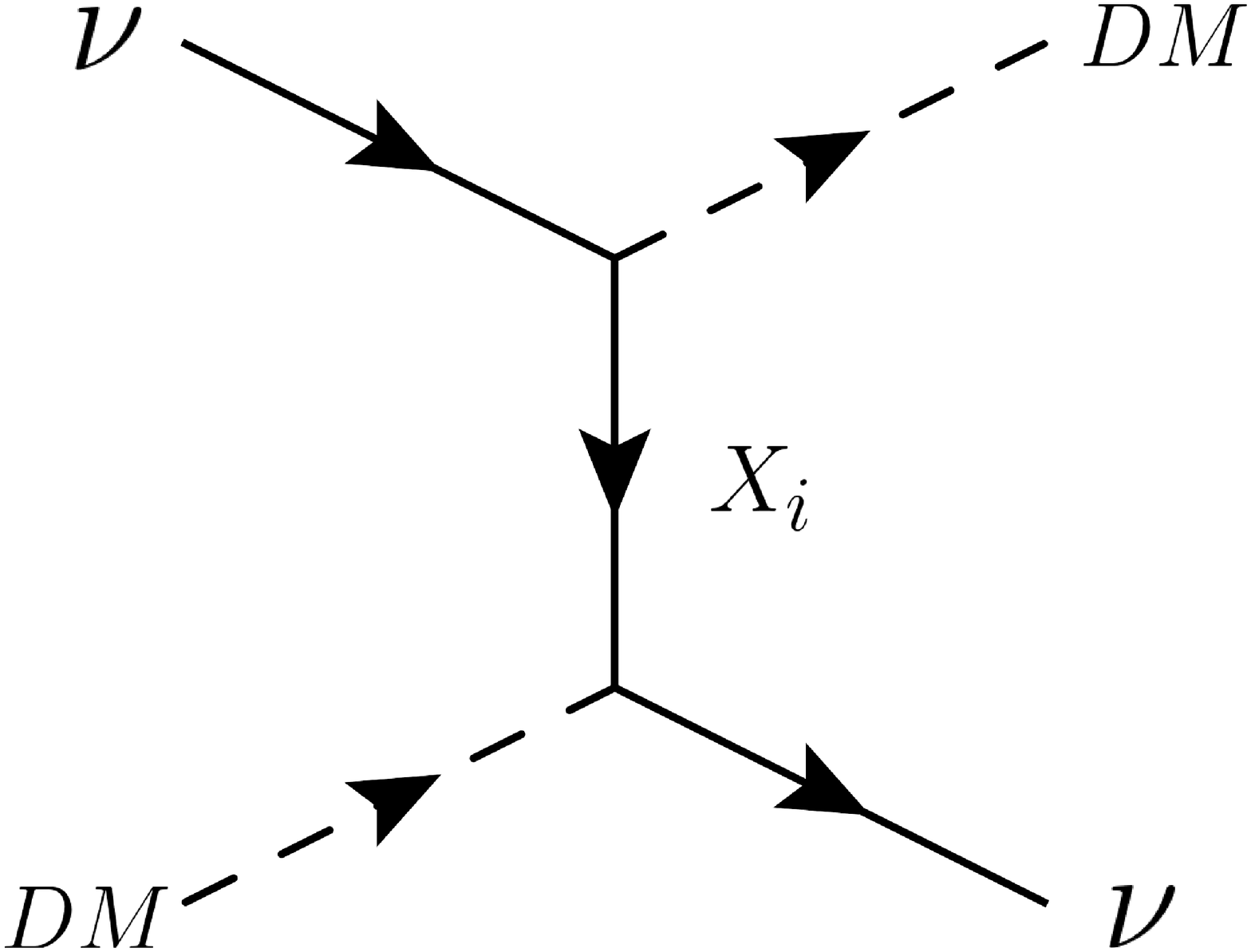}
 &
 \includegraphics[width=0.2\textwidth]{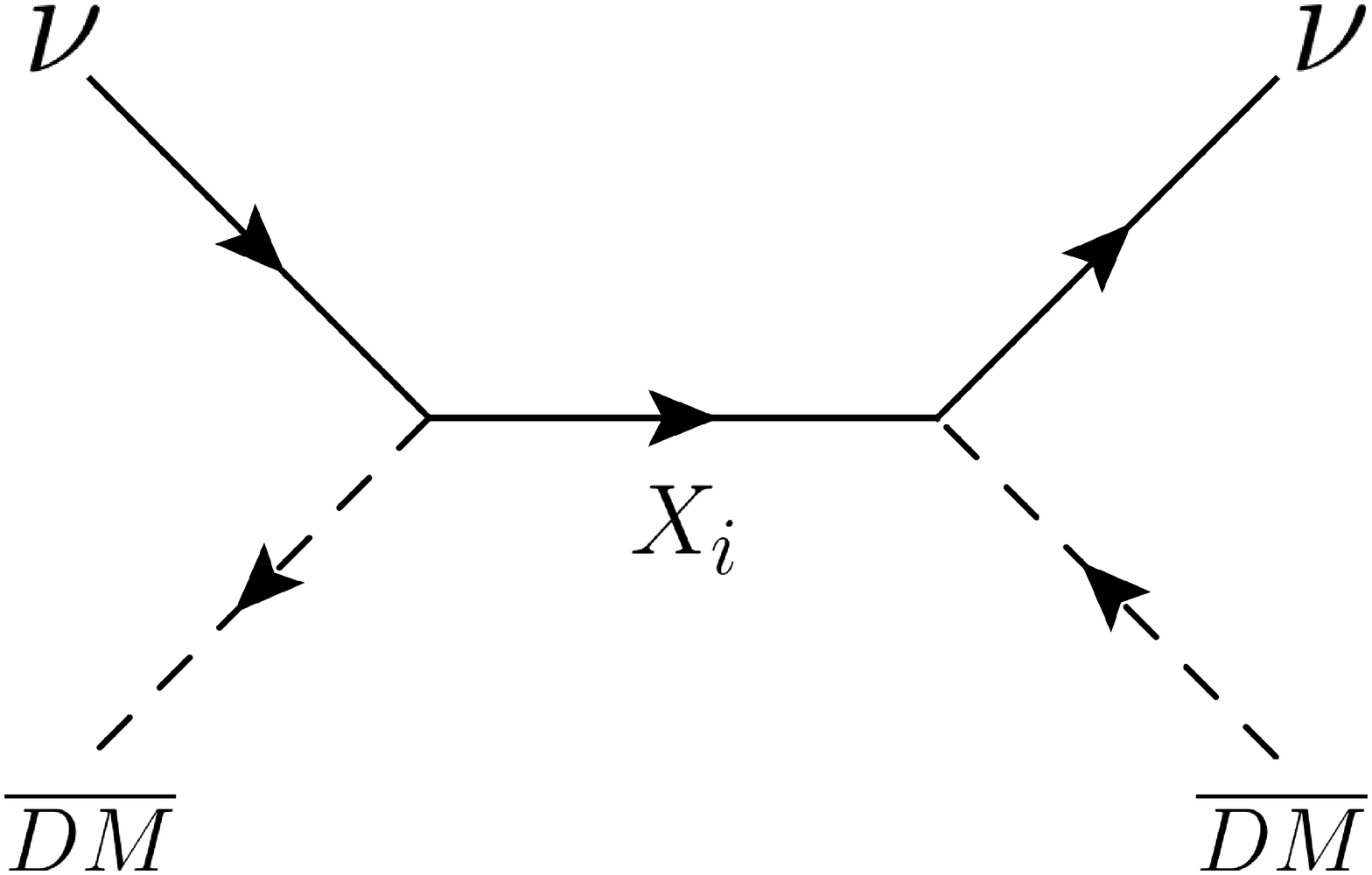}
     \end{tabular}
\end{center}
\caption{ Feynman diagrams for  the  scattering of  neutrino and complex scalar dark matter mediated by a fermion in the scenario of \eq{model3}.}
\label{diagram}
\end{figure}

In the case of the model in \eq{model3}, the calculation of the $s$ and $u$ channel diagrams in Fig.~\ref{diagram} for the forward elastic scattering gives 
\begin{eqnarray}
\Sigma_1\, (\textrm{or}\, \bar\Sigma_1) & \simeq & \frac{\lambda^{(T)}}{2} \frac{\rho_{{DM}}}{m^2_{DM}}
 \frac{\pm \epsilon\, 2m_{DM} E_\nu - m_X^2}{m_X^4-4 m^2_{DM} E^2_\nu} , \label{Sigma1} \\
\Sigma_2\, (\textrm{or}\,  \bar\Sigma_2) & \simeq & \frac{\lambda^{(T)}}{2} \frac{\rho_{DM}}{m^2_{DM}}
\frac{\pm \epsilon\, m_X^2 - 2 m_{DM} E_\nu }{m_X^4-4 m^2_{DM} E^2_\nu} ,\label{Sigma2}
\end{eqnarray}
where the coupling matrix $\lambda$ is defined by $\lambda_{\alpha\beta} \equiv g^*_{\alpha i} g_{\beta i}/2$
for the transition $\nu_\beta \rightarrow \nu_\alpha$ and 
 the same mass $m_X$ is assumed for the mediators $f_i$, and $m_{DM}=m_\phi$ is the dark matter mass.   
From the expression (\ref{Sigma1}), one can find a remarkable property that the meduim mass matrix $\tilde M$ becomes symmetric
only for the symmetric dark matter distribution ($\epsilon=0$) 
as the asymmetric medium distinquishes neutrinos ($u_L$) and anti-neutrinos ($u_R$) as shown in (\ref{uLuR}). 

Eq.~(\ref{Sigma2}) tells us that  the medium (DM)  potentials are given  by
 \begin{equation} \label{VDM}
 V^{DM}_{\nu,\bar\nu} \simeq \frac{\lambda^{(T)}}{2} \frac{\rho_{DM}}{m_{DM}}
\frac{\pm \epsilon\, m_X^2 - 2 m_{DM} E_\nu }{m_X^4-4 m^2_{DM} E^2_\nu}.
\end{equation}
This formula is applicable to the all three scenarios with different candidates of dark matter and mediators.
Notice that the case considered in~\cite{Ge:2019tdi} \footnote{The sign of the anti-neutrino potential was opposite to ours. 
But the authors agreed with our result in a private communication.} should correspond to  the limit $m_X\rightarrow 0$ in our formulation with $\epsilon=0$.
Like the medium effect to the mass matrix, the medium potentials
for neutrinos and anti-neutrinos are the same for a  symmetric medium, and receive opposite contributions
from the asymmetry ($\epsilon \neq 0$). 
This violation of CPT symmetry due to the environmental matter effect needs to be distinguished from the theory with CPT violation~\cite{Kostelecky:2003cr,Barenboim:2017ewj,Barenboim:2018lpo,Majhi:2019tfi,Ohlsson:2014cha}.

\medskip

\begin{figure}[!t]
\begin{center}
\begin{tabular}{c} 
 \includegraphics[width=0.45\textwidth]{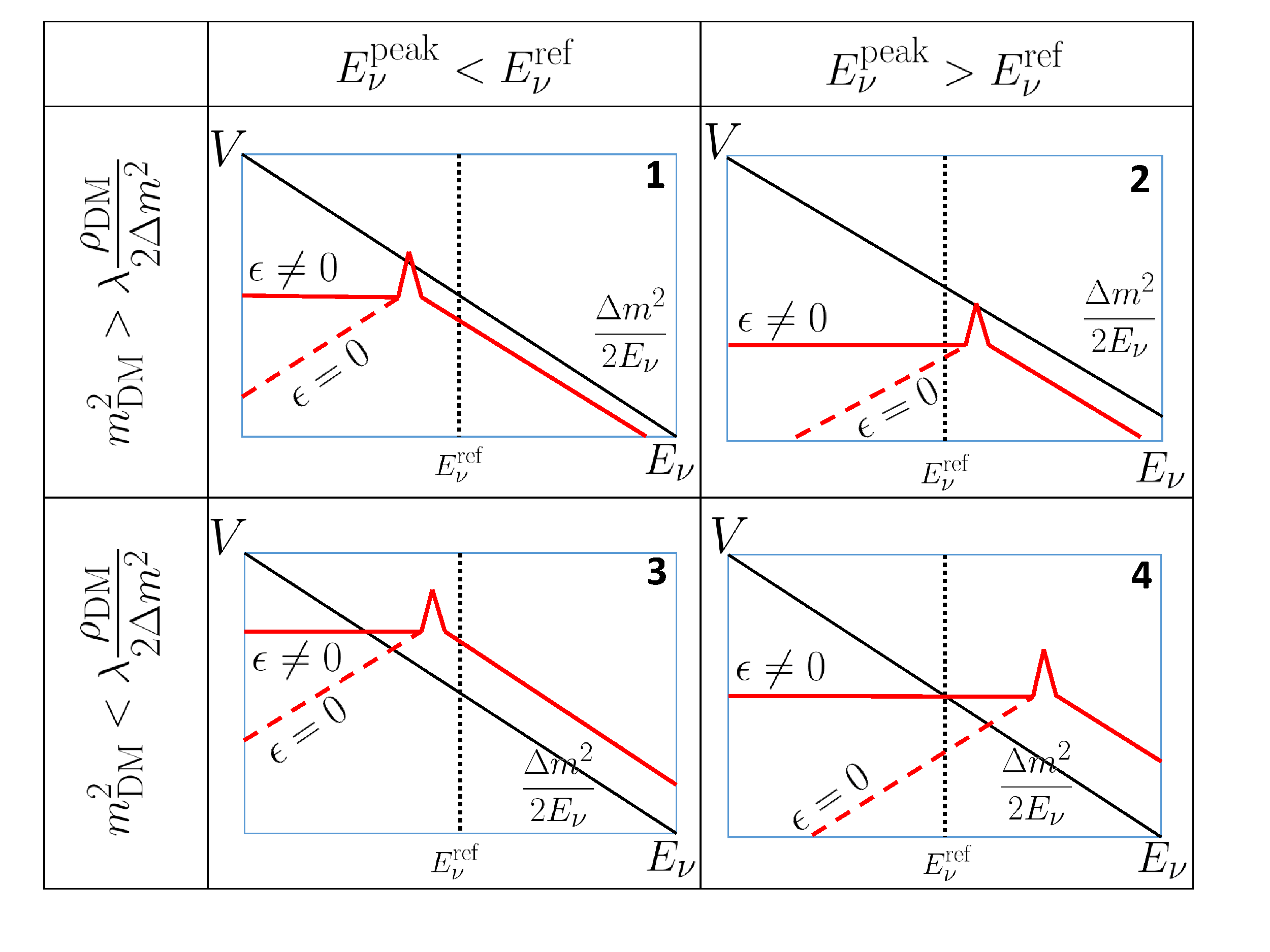}
    \end{tabular}
\end{center}
\caption{The schematic plot for the shape of the medium potential for different regimes of  $m_{DM}$ and $E_\nu^{peak}$. 
In the small boxes, we show the absolute value of the medium potential with respect to the neutrino energy. 
The solid (dashed) red line is for asymmetric (symmetric) distribution of DM. For comparison, $\Delta m^2/2E_\nu$ is shown by
the black solid line.  The black dotted vertical line denotes the reference neutrino energy scale $E_\nu^{\rm ref}$ in a certain experiment of interest. 
}
\label{mFmphi}
\end{figure}

The configurations of the medium potential (\ref{VDM}) are presented in the small boxes  of Fig.~\ref{mFmphi}.
One can consider four different regions depending on whether $E^{peak}_\nu$ defined by
\dis{ 
E^{peak}_\nu = \frac{m_X^2}{2 m_{DM}},
}
is larger or smaller than the reference neutrino energy $E_\nu^{\rm ref}$ and also whether 
the medium potential at high energy is larger or smaller than $\Delta m^2/2E_\nu$, that is,
\dis{
m_{DM}^2 >   \frac{\lambda \rho_{\rm DM}}{2|\Delta m^2|}, \quad \textrm{or} \quad m_{DM}^2 <  \frac{\lambda \rho_{\rm DM}}{2|\Delta m^2|}.
}

In the region 1 and 2, the medium potential is  sub-dominant and may give small modification to the standard oscillation, or a peculiar  feature can show up at the energy of the peak. In the region 3 and 4, if $E^{\rm peak}_\nu$ is in the range of 1 MeV --100 GeV, there would appear a high distortion in  various standard neutrino oscillation data and thus it is strongly disfavored.  
 In region 4, there is no signals at low energy data, however the future experiments of neutrino oscillation at high energy can probe this region. 
The boundary of the regions 1 and 3 (the dashed line) with $E^{\rm peak}_\nu\ll 1 \mev$ is of particular interest as the medium potential mimics the SM mass term and can explain the neutrino oscillation data even with massless neutrinos.

More specifically, 
when $|\epsilon| m^2_X \gg 2 m_{DM} E_\nu$, we get the matter potential
\begin{eqnarray} \label{VDM1}
V^{DM}_{\nu,\bar\nu} & \simeq & \pm \epsilon \frac{\lambda^{(T)}}{4} \frac{\rho_{DM}}{m_{DM}^2 E_\nu^{peak} }.
\end{eqnarray}
Given the masses as chosen above, the conventional bounds on non-standard interactions (NSI)
can be applied to each component of $\epsilon \lambda^{(T)}$.
 Normalizing $V^{DM}_{\alpha\beta}$ by $V^{SM}$, we get
the standard NSI form:
\dis{ 
\varepsilon_{\alpha\beta} \approx 0.01 \lambda_{\alpha\beta} \epsilon
\left( \frac{20\mbox{meV}}{m_{DM}} \right)^2
\bfrac{1\tev}{E_\nu^{peak}}
\left( \frac{ \rho_{\rm DM}}{0.3 \mbox{GeV\,cm$^{-3}$}}\right),
}
taking $N_e\approx 1.3\times 10^{24}/\cm^3$ for the earth mantle density. 
The values of $\varepsilon$ are constrained to be smaller than around 0.1 or 0.01~\cite{Ohlsson:2012kf,Adamson:2016yso,Ge:2016dlx,Aartsen:2017nmd}, which is applicable to the case 2 or 4 in Fig.~\ref{mFmphi}. 
Considering a rough bound of  $|\varepsilon| \lesssim 0.01$, we find the allowed region of $m_{DM}$:
\dis{
\mdm \gtrsim 20 \mbox{ meV}|\epsilon|^{1/2} |\lambda|^{1/2} \bfrac{\tev}{E^{peak}_{\nu}}^{1/2}.
}
An independent bound comes from the medium mass dressed by $\Sigma_1/\bar\Sigma_1$. Requiring  the correction is less than about  $1\%$, we get
\dis{
\mdm \gtrsim 10^{-4} \mbox{ meV} |\lambda|^{1/3} \bfrac{\tev}{E^{peak}_{\nu}}^{1/3}.
}
Given specific value of $\lambda$ and $\epsilon$, the stronger limit is to be taken.

As noted before, in the case of  $m^2_X \ll 2 m_{DM} E_\nu$, the medium potential becomes 
\begin{eqnarray} \label{VDM2}
V^{DM}_{\nu,\bar\nu} & \simeq & \frac{\lambda^{(T)}}{2} \frac{\rho_{DM}/m_{DM}^2}{2 E_\nu}, \\
&\approx&  \frac{3\times 10^{-3} \mbox{eV}^2}{2E_\nu} \lambda^{(T)} \left( {20\mbox{meV} \over m_{DM}}\right)^2,
\nonumber
\end{eqnarray}
which behaves same as the standard neutrino masses and mixing explaining the observed neutrino oscillations for $m_X \ll 200 \mbox{eV} \sqrt{(m_{DM}/20\mbox{meV})(E_\nu/1\mev) }$.  Therefore, the experimental data can be fitted by the couplings given by 
\begin{eqnarray}
\label{lambdaDM}
\lambda &=& \frac{2m^2_{DM}}{\rho_{DM}} U^* {\rm diag}(\Delta m^2) U^T ,\\
&\simeq  &
\begin{pmatrix}
0.026&0.091&0.085\\
0.091&0.381&0.408\\
0.085&0.408&0.478
\end{pmatrix}\bfrac{m_{DM}}{20\mbox{meV}}^{2}\bfrac{0.3\gev\cm^{-3}}{\rho_{DM}},
\nonumber
\end{eqnarray}
where ${\rm diag}(\Delta m^2)_{ii}=m_i^2-m_1^2 $ and $U$ is the neutrino mixing matrix 
assuming the normal hierarchy and vanishing CP phases. 
In this model of ``dark matter assisted neutrino oscillation", the leading correction with non-vanishing $\epsilon$ may help to fit better the observations of neutrinos from Sun and the reactor anti-neutirno experiments~\cite{Kopp:2010qt,Ge:2019tdi}.  
It would be interesting to observe the oscillation of high energy neutrino which might be affected by this medium effect~\cite{Lunardini:2000swa,Aartsen:2015knd,Karmakar:2018fno,Choi:2019ixb,Farzan:2018pnk,Ahlers:2018yom,Bustamante:2019sdb}.
In the future, the measurement of the absolute value of neutrino mass may rule out this model.

When the observed neutrino oscillations come from the standard neutrino mass matrix, the medium corrections are constrained.
That is, requiring the medium corrections less than about $1\%$ for the region 1 or 3, we find 
\dis{
\mdm \gtrsim& \,200\, \mbox{meV}\, |\lambda|^{1/2},\\
\mdm \gtrsim & \,0.3\,  \mbox{meV}\, |\epsilon|^{1/3} |\lambda|^{1/3} \bfrac{1\mev}{E_\nu}^{1/3},
}
from $\Sigma_2$, and $\Sigma_1$, respectively.

\medskip

{\it Apparent CPT violation}:
The Lagrangian itself  is CPT invariant, but the medium of  asymmetric  DM  is not.  
Thus, the effective neutrino mixing and mass-squared differences are modified in a different way for neutrinos and antineutrinos. 
Such a CPT violation may appear in precision measurement of neutrino oscillations, or is highly constrained by the present data, particularly if the peak energy resides between 1 MeV and 100 GeV where the standard neutrino oscillation has been confirmed.

\medskip

{\it Two-flavor oscillation}:
To see the medium effect in more detail,  let us consider  the two-flavor ($\nu_{\mu}$--$\nu_\tau$) oscillation described by 
the effective Hamiltonian: 
\begin{eqnarray}
\mathcal{H}_M &=& \mathcal{H}_{\rm vac} 
+ \begin{pmatrix}
V_{\mu\mu} & V_{\mu\tau} \\ V^*_{\mu\tau} & V_{\tau\tau}
\end{pmatrix},
\end{eqnarray}
where $\mathcal{H}_{\rm vac}$ is for the oscillation in  the standard model 
\begin{eqnarray}
\mathcal{H}_{\rm vac} &=& \frac{\Delta m^2}{4E} \begin{pmatrix}
-\cos 2\theta & \sin 2\theta \\ \sin 2\theta & \cos 2\theta
\end{pmatrix}. \label{Ham:vac}
\end{eqnarray}
Up to the diagonal term proportional to the identity matrix, which is irrelevant to the oscillation, $\mathcal{H}_M$
can be re written as
\begin{eqnarray}
\mathcal{H}_M &=& \frac{\Delta m^2}{4E} \begin{pmatrix}
-(\cos 2\theta -x) & \sin 2\theta + y \\ \sin 2\theta + y& \cos 2\theta -x
\end{pmatrix}, \label{Ham:mat}
\end{eqnarray}
where
\dis{
x\equiv  \frac{ (V_{\mu\mu}-V_{\tau\tau})/2}{\Delta m^2 / 4E},~~ \textrm{and}~~
y\equiv \frac{V_{\mu\tau}}{\Delta m^2 / 4E}. \label{xy}
}
Thus one obtains the usual  mixing angle and mass-squared difference in the medium given by
\dis{\label{mixing}
&\sin^2 2\theta_M = \frac{ (\sin 2\theta + y)^2 }{(\cos 2\theta - x)^2 + (\sin 2\theta + y)^2 }, \\
&\Delta m^2_M = \Delta m^2 \sqrt{(\cos 2\theta - x)^2 + (\sin 2\theta + y)^2 },
}
which  gives the transition probability in the medium:
\begin{eqnarray}
P_M\left(\nu_\mu \to \nu_\tau \right) = \sin^2 2\theta_M \sin^2\left(\Delta m^2_M \frac{L}{4E}\right).
\end{eqnarray}

\begin{figure}[!t]
\begin{center}
\begin{tabular}{c} 
 \includegraphics[width=0.45\textwidth]{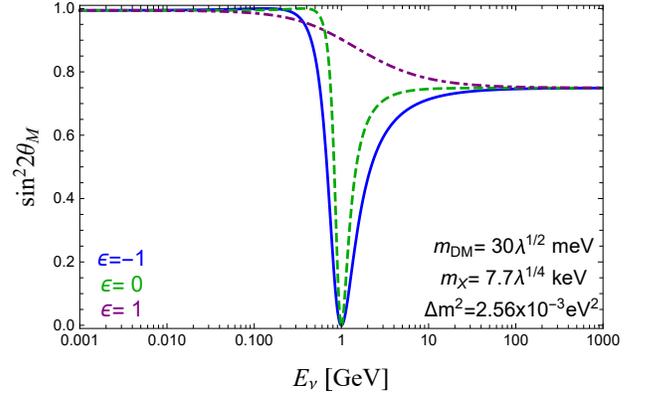}
   \end{tabular}
\end{center}
\caption{The plot of $\sin^22\theta_M$ of neutrinos as a function of $E_\nu$ with non-vanishing $\lambda_{\mu\mu}=\lambda$.
The solid, dashed, and dash-dotted lines are for the DM asymmetry $\epsilon=-1,0$, and 1, respectively.
Here we used $\Delta m^2= 2.56\times 10^{-3} \ev^2$.}  
\label{sin2}
\end{figure}

In Fig.~\ref{sin2}, we show the change of $\sin^22\theta_M$ for neutrinos in terms of $E_\nu$ 
taking only one non-vanishing coupling 
$\lambda=\lambda_{\mu\mu}$, and the masses of  $m_{DM}= 30\sqrt{\lambda}\, \miliev$ and
 $m_X=7.7\sqrt[4]{\lambda}\,\kev$.  
The solid, dashed, and dash-dotted lines correspond to the DM asymmetry   $\epsilon=-1,0$, and 1, respectively  for neutrinos. For anti-neutrinos, one can just take the opposite sign of  $\epsilon$.
Our parameters are chosen to get $E^{peak}_\nu=1\gev$ and $y=0$ and $x\rightarrow 0.75$ at the high energy limit.
At low energy, $\sin^22\theta_M$ shows the standard mixing since the medium effect is subdominant. 
As the neutrino energy approaches the peak value $1\gev$, the DM potential enhances dramatically leading to $x\gg1$, and thus 
the mixing goes to zero. Thus  $\sin^22\theta_M$ becomes negligible and $\Delta m_M^2$ is high,
and thus the transition probability becomes low at this point.  At higher energy, $V^{\rm DM}\propto 1/2E_\nu $ is comparable to the standard oscillation $\Delta m^2/2E_\nu$ and thus  new mixing angle and mass-squared difference is obtained.

\begin{figure}[!t]
\begin{center}
\begin{tabular}{c} 
 \includegraphics[width=0.45\textwidth]{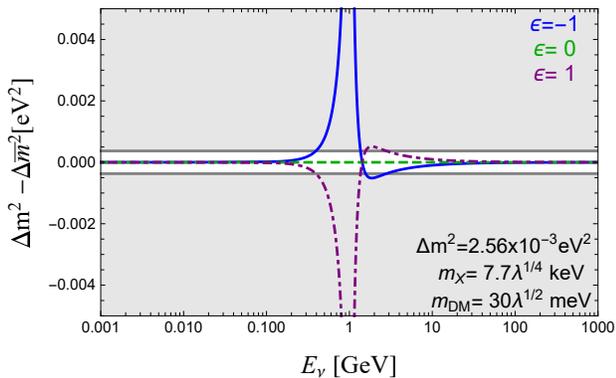}
   \end{tabular}
\end{center}
\caption{  The plot of  $\Delta m^2_M - \Delta\bar{m}^2_M$ for different DM asymmetry $\epsilon=\pm1$ (solid, dash-dotted). 
Obviously, no difference appears for $\epsilon=0$ (dashed). }
\label{CPT}
\end{figure}

Fig.~\ref{CPT} shows $\Delta m_M^2 - \Delta\bar{m}^2_M$ for the same parameters as in Fig.~\ref{sin2}. 
The resonance peak should be taken with caution as we ignored the width effect 
which will be tiny for small couplings $|\lambda| \ll 1$.
Around the peak energy, the difference is amplified for non-zero DM asymmetry, while the effect becomes moderate away from the peak energy. 
The grey region is excluded by the measurement of the $\Delta m^2$ difference between neutrinos and anti-neutrinos~\cite{Barenboim:2017ewj}
\dis{\label{CPTbound}
&|\Delta m_{21}^2 - \Delta\bar{m}^2_{21}|<4.7\times 10^{-5} \ev^2,\\
&|\Delta m_{31}^2 - \Delta\bar{m}^2_{31}|<3.7\times 10^{-4}\ev^2,
}
where the second bound is applied in the plot.

\begin{figure}[!t]
\begin{center}
\begin{tabular}{c} 
 \includegraphics[width=0.45\textwidth]{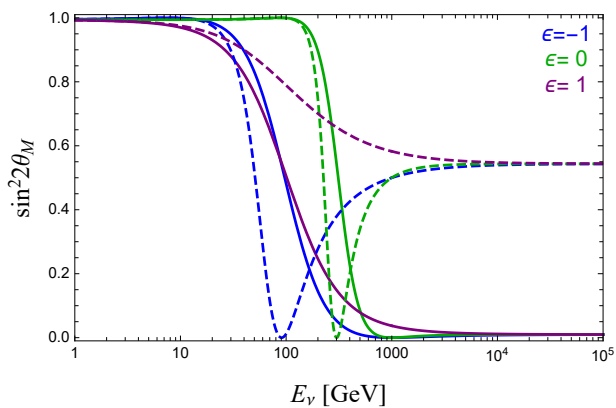}
   \end{tabular}
\end{center}
\caption{ The same plot as Fig.~\ref{sin2} but with  $E^{peak}_\nu= 1\tev$: the solid and dashed lines are for  $\lambda_{\mu\mu}\neq0$ with $x\rightarrow10, y=0$, and  $\lambda_{\mu\mu}=\lambda_{\mu\tau}\neq 0$  with $x\rightarrow10, y\rightarrow 10$  at high energy limit, respectively.}
\label{sin22}
\end{figure}

In Fig.~\ref{sin22}, we show the same plot as Fig.~\ref{sin2} but with different parameters which gives $E^{peak}_\nu= 1\tev$ and $x\rightarrow10, y=0$ (solid) and $x\rightarrow10, y\rightarrow 10$ (dashed) at the high energy limit. 
This scenario (the boundary of the region 2 and 4) shows an interesting possibility that the low energy oscillations come from the standard neutrino mass term, but the high energy oscillations from the medium effect. As can be seen in the plot, the high energy oscillation parameters are controlled by the flavor structure of the medium potential.

\medskip

{\it Conclusion and Discussion}: We provided a systematic study of neutrino oscillations in a medium of dark matter which generalizes the SM matter effect. A general formula is derived to describe the medium effect in various scenarios of dark matter and its mediator to neutrinos.  Apparent CPT violation arises from the asymmetric distribution of DM which distinguishes neutrinos and anti-neutrinos. 
Thus precise determination of the neutrino oscillation parameters may be able to reveal the presence of the DM asymmetry. 
The medium potential has a resonance peak at $E_\nu = m_X^2/2 m_{DM}$ which should be below 1 MeV or above 100 GeV not to 
spoil the standard oscillation picture. 

In the former case, the medium potential mimics the standard oscillation parameters and thus solar and atmospheric neutrino data might be accounted for even with massless neutrinos.
This  ``dark matter assisted neutrino oscillation" could be a good alternative to the standard oscillation paradigm if the absolute neutrino mass measured in neutrinoless beta decay, single beta decay or cosmological observations turns out to be unexpectedly small~\cite{Smirnov:2016xzf}.
In the latter case, ultra-high energy neutrino oscillations are described by the symmetric medium effect, and thus could be totally different from the standard neutrino oscillations which have been confirmed by various experiments at lower energies.

Our formulation brings many interesting questions: what will be the implications to the standard neutrino oscillations; 
how our medium parameters are constrained by various cosmological and astrophysical observations; and  how a low-energy scenario for the dark sector coupling to neutrinos can arise from a UV-completed theory \cite{future}.

\medskip

{\it Acknowledgments}.
The authors thank  Alexei Smirnov, Shao-Feng Ge, and Carlos Arguelles for useful discussions.
K.-Y.C. was supported by the National Research Foundation of Korea(NRF) grant funded by the Korea government(MEST) (NRF-2019R1A2B5B01070181).   E.J.C. acknowledges support  from InvisiblesPlus RISE No. 690575. 
\medskip



\end{document}